\documentclass{PoS}

\usepackage{graphics}
\usepackage{amsmath}
\newcommand{\be}{\begin{eqnarray}}
\newcommand{\ee}{\end{eqnarray}}
\newcommand{\vect}{\overrightarrow}

\newcommand{\benum}{\begin{enumerate}}
\newcommand{\eenum}{\end{enumerate}}

\newcommand{\bitem}{\begin{itemize}}
\newcommand{\eitem}{\end{itemize}}

\title{Semileptonic $D$-decays and Lattice QCD}

\ShortTitle{Semileptonic $D$-decays}

\author{Damir Be\'cirevi\'c, \speaker{Benjamin Haas} \\
	Laboratoire de Physique Th\'eorique (Bat. 210)\\
	Universit\'e Paris-Sud, Centre d'Orsay \\
	F-91405 Orsay-Cedex, France\\
        E-mail: \email{Benjamin.Haas@th.u-psud.fr}}

\author{Federico Mescia\\
        INFN, Laboratori Nazionali di Frascati,\\
	Via E. Fermi,40 \\
	I-00044 Frascati (Rome) Italy \\}

\abstract{We explore four different strategies to extract the $D$-meson semileptonic decay 
form factors from the Green functions computed in QCD numerically on the lattice.
From our numerical tests we find that two such strategies, based on the use of double 
ratios of $3$-point correlation functions, lead to an appreciable reduction of systematic 
uncertainties. This is an important step in reducing the overall uncertainty in the 
lattice QCD results for the $D$-decay form factors which are needed to determine 
the CKM entries $\vert V_{cd}\vert$ and $\vert V_{cs}\vert$ experimentally, that  are nowadays
known by imposing the unitarity of the CKM matrix.
}

\FullConference{The XXV International Symposium on Lattice Field Theory\\
		 July 30-4 August 2007\\
		 Regensburg, Germany}

\begin{document}
\bibliographystyle{plain}

\section{Introduction}
An accurate determination of the Cabibbo--Kobayashi--Maskawa (CKM) matrix elements, 
$V_{ij}$, is an  essential step in testing the Standard Model (SM). 
Like the quark masses, the couplings $V_{ij}$ are free parameters of the SM 
and therefore cannot be predicted. Instead they are extracted after 
confronting the experimental measurements to the SM theoretical expressions. 
The simplest processes in that respect are the leptonic and semileptonic decays 
of pseudoscalar mesons. In this note we consider $D$ decays, namely 
\begin{eqnarray}
 \frac{d\Gamma}{dq^2}(D_q\rightarrow P_{qq} \ell \nu_\ell)   & =&|V_{cq}|^2 \frac{G_F^2}{192 \pi^2m_{D_q}^3}\lambda^{3/2}(q^2) | F_+(q^2) |^2 \,, \cr
\label{eqn_semi}
 \Gamma(D^+_{q}\rightarrow \ell \nu_\ell) &  =& |V_{cq}|^2  \frac{G_F^2}{8\pi}f_{D_{q}}^2 m_{D_q}m_\mu^2 \left(1-\frac{m_\mu^2}{m_{D_q}} \right)\,,
\label{eqn_lept}
\end{eqnarray}
where $\ell$ is $\mu$ or $e$. The left-hand-side in the above expressions
 is measured experimentally, while the computation of hadronic form factor, $F_+(q^2)$, 
 and/or  the meson decay constant, $f_{D_{q}}$, requires a first principle description of  
 non-perturbative QCD effects. We now restrain our attention to $F_+(q^2)$, one of the two 
 form factors which parameterise the SM weak matrix element $\langle \pi(\vec{k}) \vert  (V-A)_{\mu}  \vert D(\vec{p}) \rangle 
 \equiv \langle \pi(\vec{k}) \vert  V_{\mu}  \vert D(\vec{p}) \rangle$, i.e.,
\begin{eqnarray} \label{eqn_matel}
\langle \pi(\vec{k}) \vert  V_{\mu}  \vert D(\vec{p}) \rangle = \left(p+k-q\frac{m_D^2-m_\pi^2}{q^2}\right)_\mu F_+(q^2) + q_\mu \frac{m_D^2-m_\pi^2}{q^2}F_0(q^2) \,,
\end{eqnarray}
both depending on $q^2=(p-k)^2$ only, with $q^2 \in (0,(m_D-m_\pi)^2]$. 

Lattice QCD is the only currently available method which allows us to compute this matrix element without introducing 
any extra parameter and, at least in principle, with an accuracy that can be matched to the experimental one. 
In practice, however, there is still quite a room for improvement on systematic errors. 
Here we want to address those that arise from the extraction of the matrix element~(\ref{eqn_matel}) from 
the correlation functions.
 \subsection{An abridged description of the standard procedure}
\noindent
The standard method consists in computing the $2$- and $3$-point functions, namely, 
\begin{eqnarray}
C^{\pi\pi}_2(\vec{k},t)&=&\sum_{\vec{x}}\langle (\bar{q}\gamma_5 q)_{\vec{x},t}
(\bar{q}\gamma_5 q)_{\vec{0},0} e^{i\vec{k}\vec{x}}  \rangle
 \stackrel{t \gg 0}{\longrightarrow}  \frac{{\cal{Z}}_\pi}{2E_\pi}e^{-E_\pi t} \,,\cr
C^{DD}_2(\vec{p},t)&=&\sum_{\vec{x}}\langle (\bar{c}\gamma_5 q)_{\vec{x},t} (\bar{q}\gamma_5 c)_{\vec{0},0} e^{i\vec{p}\vec{x}} \rangle
 \stackrel{t \gg 0}{\longrightarrow}  \frac{{\cal{Z}}_D}{2E_D}e^{-E_D t} \,,\cr
C^{\pi V_\mu D}_3(\vec{k},\vec{q};t,t_{source})&=&\sum_{\vect{x},\vect{z}} 
\langle (\bar{q}\gamma_5 q)_{\vec{x},t_{source}}  
(\bar{q}\gamma_\mu c)_{\vec{z},t} (\bar{c}\gamma_\mu q) _{\vec{0},0} \rangle e^{-i(\vec{q}\vec{z}-\vec{k}\vec{x})} \cr
&&\stackrel{0 \ll t \ll t_{source}}{\longrightarrow}  \frac{ {{\cal Z} _\pi}^{1/2}}{2E_\pi}e^{-E_\pi t} \langle\pi(\vec{k}) \vert  V_{\mu}  \vert D(\vec{p}) \rangle \frac{ {{\cal Z} _D}^{1/2}}{2E_D}e^{-E_D(t_{source}-t)} \,,
\end{eqnarray}
where we also indicate their asymptotic behavior, using the standard notation, 
${\cal Z}_D=\left|\langle 0 | \bar{c}\gamma_5 q | D(\vec p) \rangle \right| ^2$, 
and similar for  ${\cal Z}_\pi$. The matrix element~(\ref{eqn_matel}) corresponds to a plateau of the ratio
\begin{equation}
R=\frac{C^{\pi \gamma_\mu D}_3(q,t;t_{source})}{C^{DD}_2(\vec{k},t_{source}-t){C^{\pi\pi}_2(\vec{p},t)}}\times 
\sqrt{{\cal{Z}}_\pi}\sqrt{{\cal{Z}}_D} \stackrel{0 \ll t \ll t_{source}}{\longrightarrow}   \langle \pi(\vec{k}) \vert  V_{\mu}  \vert D(\vec{p}) \rangle\,.
\end{equation}
In case of $D$-decays the plateaus are known not to be long enough to guarantee a percent accuracy (especially when the momenta are given to 
either of the two mesons). 
Furthermore, $\sqrt{{\cal{Z}}_\pi}$ and $\sqrt{{\cal{Z}}_D}$ should be computed from a 
separate study of the $2$-point functions, the errors of which are carried over to the ratio $R$. 
Finally, a non-negligible statistical error is introduced by the multiplicative renormalization 
of the vector current, especially when consistently implementing the ${\cal{O}}(a)$ improvement of the 
Wilson quark operators on the realistic lattices and one of the quark being charmed (heavy). 
All these difficulties can be avoided by considering various 
double ratios of $3$-point correlation functions. The efficiency of the double ratios was 
introduced and tested first in {\it heavy-to-heavy}~\cite{Hashimoto:1999yp} and then 
in {\it light-to-light} decays~\cite{spqr}. In $D$-decays to a light meson an extra problem 
is related to the available kinematics on the lattice. More specifically, with the periodic 
boundary conditions the minimal momentum on realistic lattices is $(2\pi)/L$, which is too large ($L=N_La$ not large enough) 
if one is to keep the lattice spacing, $a$, sufficiently small in order to accommodate the charm quark 
mass. To get around this problem, we adopted the twisted boundary conditions (twBC) recently proposed 
in ref.~\cite{Bedaque:2004kc}. In such a way the momenta $\vec{p}$ and/or $\vec{k}$  
become $\vec{\theta}/L=(\theta_0,\theta_0,\theta_0)/L$, where the components $\theta_0$ are chosen anywhere between 
$0 \leq \theta_0 < \pi$. In this way we are able to compute the form factor $F_+(q^2)$ at several $q^2 >0$.

\begin{figure}
\begin{center}
\includegraphics[width=6cm,clip]{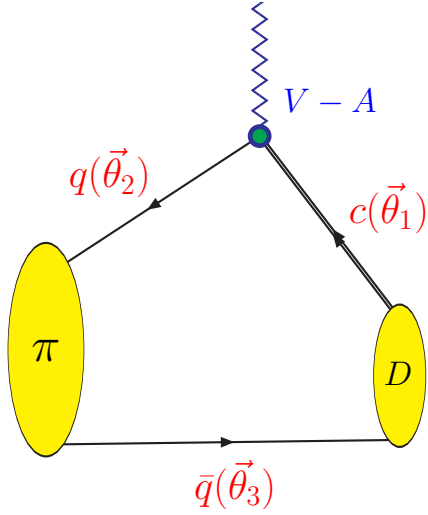}
\caption{The valence quark diagram of the $3$-point function, $C_3^{\pi V_\mu D}(\vec{k},\vec{q};t,t_{source})$. We indicate the 
twisting angles, $\vec \theta_{1,2,3}$, which are discussed in the text. }
\label{fig_3pts}
\end{center}
\end{figure}

\section{Double ratios}
In what follows we consider $4$ different strategies to increase the accuracy of extraction 
the form factor $F_+(q^2)$ for several values of $q^2$ by combining 
the $3$-point functions in suitable  double ratios. At this point we should emphasize 
that in each of the strategies discussed in this section, the multiplicative renormalization 
factors, as well as the source terms, ${\cal Z}_{\pi,D}$, cancel out. We then numerically test each of the
proposed strategies to check whether or not a plateau region is pronounced enough and 
the statistical quality of the signal satisfactory to reach a percent accuracy of the extracted form factor.

\subsection{First strategy}
\begin{figure}
\hspace{3cm}
\resizebox{8cm}{8cm}{\includegraphics{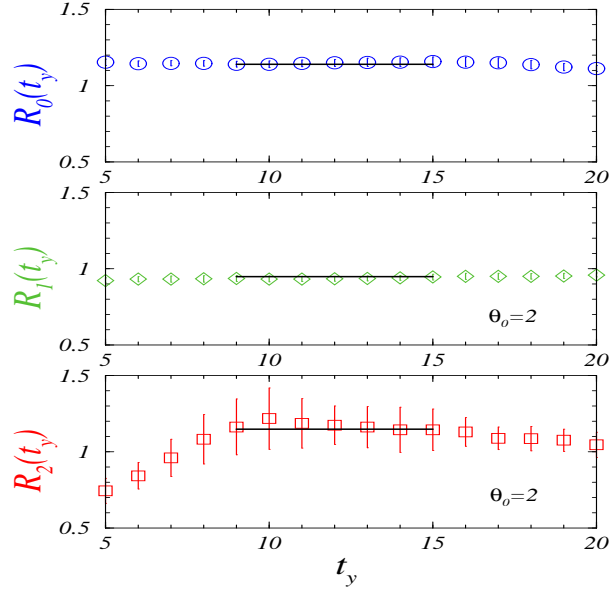}} 
\caption{Double ratios, $R_{0,1,2}(t)$, illustrating the first strategy. The lattice data refer to the simulation with Wilson ${\cal O}(a)$-improved 
quarks with  $N_F=2$ ($\beta=5.29$) and $\kappa_q=\kappa_{\rm sea}=0.1355$. The shown signals refer to $q^2\approx 1~{\rm GeV}^2$. }
\label{fig_rat012}
\end{figure}

We first keep the $D$-meson at rest and inject momenta to the pion only. 
This is done by imposing $ \vec{\theta_1}=\vec{\theta_3}=\vec{0}$ (c.f. fig.~\ref{fig_3pts}), 
while for $\vec{\theta_2}$ we choose several different values to explore the kinematics available from this decay, 
$0\leq q^2 \leq q^2_{max}$.
Similar to what has been proposed by JLQCD in their study of the $K_{\ell 3}$-decay~\cite{jlqcd}, 
we consider the following double ratios ($\vec{k}=\vec{\theta_2}/L$):
\be
\frac{C_3^{\pi V_0 D}\left(\vec{0},t\right)C_3^{DV_0 \pi}\left(\vec{0},t\right)}{C_3^{\pi V_0 \pi}\left(\vec{0},t\right)C_3^{DV_0 D}\left(\vec{0},t\right)} &\stackrel{\rm plateau}{\longrightarrow}& R_0 \,,\\
\frac{C_3^{\pi V_0 D}\left(\vec{k},t\right)C_2^{\pi\pi}\left(\vec{0},t\right)}{C_3^{\pi V_0 D}\left(\vec{0},t\right)C_2^{\pi\pi}\left(\vec{k},t\right)}  &\stackrel{\rm plateau}{\longrightarrow} &R_1\,,\\
\frac{C_3^{\pi V_i D}\left(\vec{k},t\right)C_3^{\pi V_0\pi}\left(\vec{k},t\right)}{C_3^{\pi V_0 D}\left(\vec{k},t\right)C_3^{\pi V_i\pi}\left(\vec{k},t\right)} & \stackrel{\rm plateau}{\longrightarrow}& R_2\,.
\ee
These ratios then can be cast into the expressions leading to $F_+(q^2)$, i.e., 
\be
F_+(q^2)            & = & R_1 \times  F_0(q^2_{max}) \times 
\frac{m_D+m_\pi}{m_D+E_\pi}\left[1+\frac{m_D-E_\pi}{m_D+E_\pi} \times\xi(q^2) \right]^{-1}  \label{eqn_St1.2}\,,\\
{\rm where}\quad  F_0(q^2_{max}) & = & \frac{2\sqrt{m_D m_\pi}}{m_D+m_\pi} \sqrt{R_0} \,,\\
{\rm and}\quad  \xi(q^2)       & = &1-\frac{2 m_D R_2}{(m_\pi+E_\pi)+(m_D-E_\pi ) R_2}  \label{eqn_St1.3}\,.
\ee
For short we wrote $q^2 \xi(q^2)=(m_D^2 - m_\pi^2)[F_0(q^2)/F_+(q^2) - 1]$. 
The quality of the plateaus is presented in fig.~\ref{fig_rat012}. To that end we use the publicly available 
ensembles of the SU(3) gauge field configurations, produced by the QCDSF collaboration by using 
the ${\cal{O}}(a)$-improved Wilson quark action with $N_F=2$~\cite{Brommel:2006ww}. We computed 
the quark propagators and correlation functions on the configurations gathered at $\beta=5.29$, 
corresponding to $a\simeq 0.08~{\rm fm}$, on the $24^3\times48$ lattice, each time keeping the light valence quark mass
equal to that of the sea quark. 
In fig.~\ref{fig_rat012} we see that the signals for $R_0$ and $R_1$ are indeed very good. 
The fact that the signal for $R_2$ is not as good does not trouble the whole strategy because 
it is only needed to compute $\xi(q^2)$, which itself is a correction to $1$ in both eq.~(\ref{eqn_St1.3}) and in eq.~(\ref{eqn_St1.2}). 
On the other hand the fact that the quality is less good for $R_2(t)$ is expected as its computation involves the 
correlation function with with ``$\gamma_i$"-matrices (mixing the ``large" and ``small" components of the Dirac
spinors), in contrast to $R_{0,1}(t)$ in which we compute the correlation 
functions with  ``$\gamma_0$"-matrix only.

\subsection{Second and third strategies}
\begin{figure}
\begin{center}
\hspace{-1cm}\resizebox{8cm}{8cm}{\includegraphics{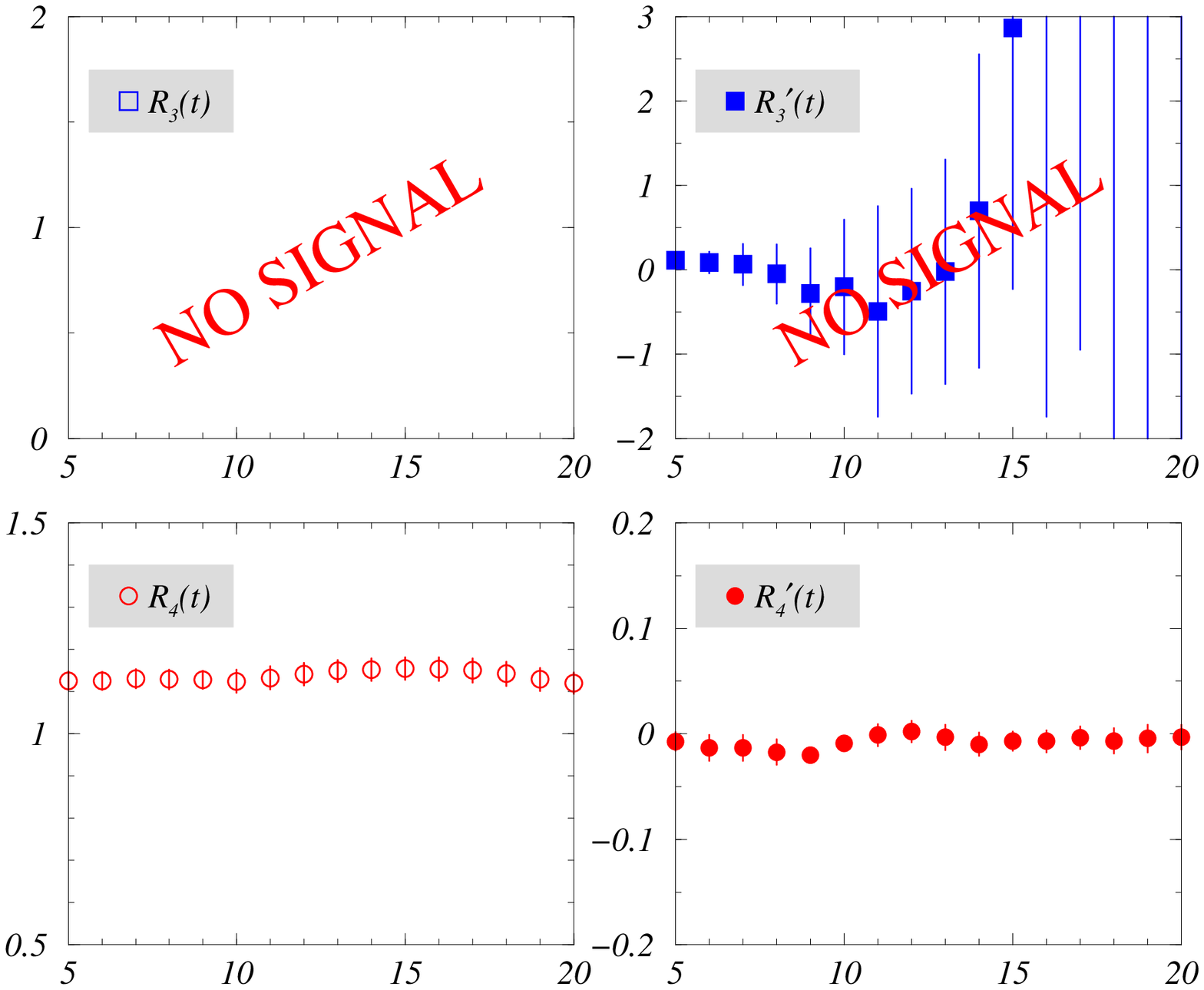}}\hspace{1.5cm}
\resizebox{6cm}{8cm}{\includegraphics{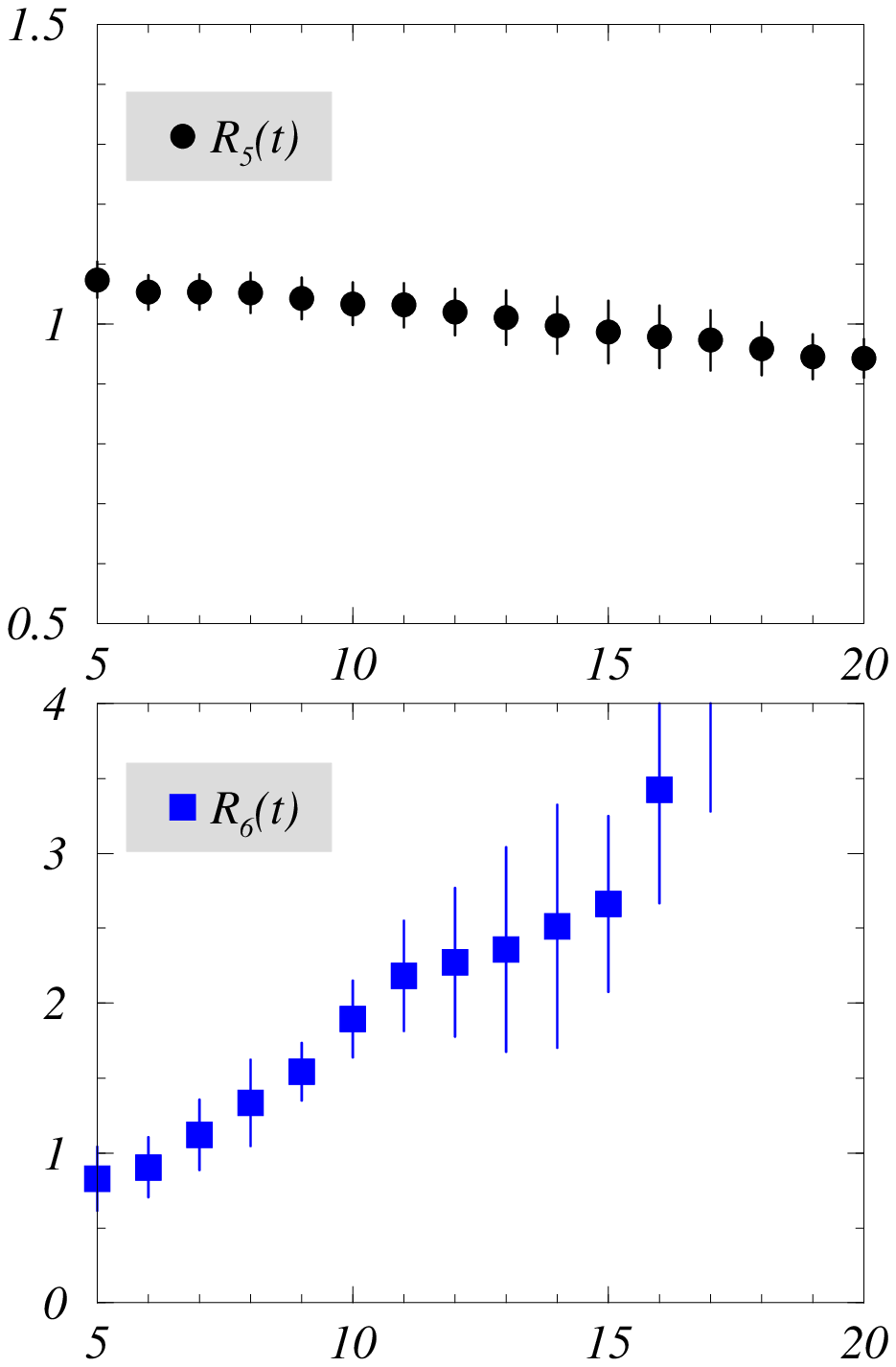}}\vspace{-0.5cm}
\end{center}
\caption{Illustration of the signals for 
the double ratios used in the second, 
third  ($R_{3}$, $R_{4}$,  $R_{3}^\prime$,  $R_{4}^\prime$) and fourth strategies ($R_{5}$, $R_{6}$),  defined in the text.}
\label{fig_rat34}
\end{figure}

Next we consider the kinematics in which we either keep the $D$-meson at rest 
($ \vec{\theta_1}=\vec{\theta_3}=\vec{0}$) and inject momenta to the pion source 
($\vec{\theta_2}=\vec{\theta_q}$), or keep the pion at rest ($\vec{\theta_1}=\vec{\theta_3}=\vec{0}$) 
and inject momenta to $D$-meson ($\vec{\theta_1}=\vec{\theta_c}$). Notice that keeping the $q^2$ fixed requires  that
$\vec{\theta_c}=(m_D/m_\pi)\times \vec{\theta_q}$. 
We then build the following four double ratios:
\be\label{strat23}
\frac{C_3^{\pi V_0 D}\left(\vec{0},\vec{p},t\right)C_3^{DV_0 \pi}\left(\vec{p},\vec{0},t\right)}{C_3^{\pi V_0 \pi}\left(\vec{0},
\vec{0},t\right)C_3^{DV_0 D}\left(\vec{p},\vec{p},t\right)}\stackrel{\rm plateau}{\longrightarrow} R_3\,,\quad\quad
\frac{C_3^{DV_0 \pi}\left(\vec{0},\vec{k},t\right)C_3^{\pi V_0 D}\left(\vec{k},\vec{0},t\right)}{C_3^{\pi V_0 \pi}
\left(\vec{k},\vec{k},t\right)C_3^{DV_0 D}\left(\vec{0},\vec{0},t\right)}\stackrel{\rm plateau}{\longrightarrow} R_4 \,,\cr
\frac{C_3^{\pi V_i  D}\left(\vec{0},\vec{p},t\right)C_3^{D V_i 
 \pi}\left(\vec{p},\vec{0},t\right)}{C_3^{\pi V_0 \pi}\left(\vec{0},\vec{0},
 t\right)C_3^{D V_i D}\left(\vec{p},\vec{p},t\right)}\stackrel{\rm 
 plateau}{\longrightarrow} R_3^\prime\,,\quad\quad\frac{C_3^{D V_i  \pi}\left(\vec{0},
 \vec{k},t\right)C_3^{\pi V_i D}\left(\vec{k},\vec{0},t\right)}{C_3^{\pi V_i 
  \pi}\left(\vec{k},\vec{k},t\right)C_3^{DV_0 D}\left(\vec{0},\vec{0},t\right)}\stackrel{\rm plateau}{\longrightarrow} R_4^\prime\,,
\ee
where $\vec{p}=\vec{\theta_1}/L$, and $\vec{k}=\vec{\theta_2}/L$. 
In terms of form factors, the above ratios read
\begin{equation}
\hspace{-0.5cm}
\nonumber
R_3=\frac{\left[E_D+m_\pi+(E_D-m_\pi)\xi(q^2)\right]^2}{2m_\pi2E_D}[F_+(q^2)]^2\,,
\quad R_4=\frac{\left[m_D+E_\pi+(m_D-E_\pi)\xi(q^2)\right]^2}{2E_\pi 2 m_D} [F_+(q^2)]^2 \,,
\end{equation}
\begin{equation}
\hspace{-3.39cm}
R_3^{\prime}=p_i\frac{1-\xi^2(q^2)}{4 m_\pi} [F_+(q^2)]^2\,, \hspace{3.24cm}R_4^{\prime}=k_i\frac{1-\xi^2(q^2)}{4 m_D} [F_+(q^2)]^2 \,.
\end{equation}
Note that UKQCD~\cite{Boyle:2007wg} recently considered $R_3$ and $R_3^\prime$  to probe $F_+^{K\to \pi}(0)$. 
We now need to combine  two of the above four double ratios to obtain $F_+(q^2)$, i.e.,  either 
$R_3$ with $R_3^\prime$, $R_4$ with $R_4^\prime$, $R_3$ with $R_4^\prime$, or $R_3^\prime$ with $R_4$. 
In fig.~\ref{fig_rat34}  we show the quality of the signals corresponding to the same set-up as 
the signals displayed in fig.~\ref{fig_rat012}. We observe that only $R_4$ and $R_4^\prime$ are reasonably good, 
whereas  $R_3(t)$ and $R_3^\prime(t)$ do not exhibit plateaus, likely due to the fact that $m_D/m_\pi$ is large, 
so that $\vert\vec{\theta_c}\vert$ in $\vec{\theta_c}=(m_D/m_\pi)\times \vec{\theta_q}$ is such that the twBC simply destroys the signal. 
Therefore, our tests suggest that the ratios $R_4$ and $R_4^\prime$  can be used to compute the form factor $F_+(q^2)$ 
for various values of $q^2$. 
 
\subsection{Fourth strategy}
If one wants to study the shape of the form factor (i.e., its $q^2$-dependence), then it is worth trying to 
impose the twBC on the spectator ($\vec{\theta_3}\neq 0$) without twisting the other two quarks 
($\vec{\theta_1}=\vec{\theta_2}=\vec{0}$). In such a way we may probe many values of $q^2$ but 
never reach $q^2=0$, because in this set-up the condition $E_D=E_\pi$ does not allow a real 
solution in $\vert \vec p\vert =\vert \vec k\vert =\vec \theta_3/L$. 
To test this option we  consider the following two ratios:
\begin{equation}\label{strat4}
\frac{{C_{3\theta}^{\pi V_0 D}}\left(t\right)  C_{3\theta}^{DV_0 \pi}\left(t\right)}{C_{3\theta}^{\pi 
V_0 \pi}\left(t\right)C_{3\theta}^{DV_0 D}\left(t\right)} \stackrel{\rm plateau}{\longrightarrow} R_5 \,,\quad\quad
\frac{C_{3\theta}^{D V_i  \pi}\left(t\right)C_{3\theta}^{\pi V_i 
D}\left(t\right)}{C_3^{\pi V_i  \pi}\left(t\right)C_3^{D V_i 
 D}\left(t\right)}\stackrel{\rm plateau}{\longrightarrow}  R_6 \,,
\end{equation}
where index ``$\theta$" is used to distinguish that the spectator quark is actually twisted. 
Expressed in terms of  form factors 
\be
R_5=\frac{\left[E_D+E_\pi+(E_D-E_\pi)\xi(q^2)\right]^2}{2E_D2E_\pi} [F_+(q^2)]^2 \,,\quad \quad
R_6=[ F_+(q^2)]^2\,.
\ee
The corresponding signals from our numerical study are shown in fig.~\ref{fig_rat34}. 
While the statistical quality of $R_5(t)$ is reasonably good, the signal for $R_6(t)$ is not 
promising if we are after a strategy that could lead us to a percent
accuracy on the extracted form factor value. 
Although we did not try it, we suspect the flatness of $R_5$ could be achieved by a judicious choice of smearing. 
We point out, however, that our numerical evaluation of the ratios~(\ref{strat4}) indicate the large statistical 
errors so that this strategy is not competitive with the first or the third ones discussed in this section.

\section{Summary}
It this note we report on the results of our exploratory study in which we use various double ratios 
and twisted boundary condition on the quark propagators in order to extract the form factor relevant to
the semileptonic heavy-to-light $D$-decays to a percent accuracy. In total we proposed four different strategies, 
which we then tested numerically on the set of unquenched ($N_F=2$) 
gauge field configurations. On the basis 
of our analysis we conclude that the first (double ratios $R_{0,1,2}$) and the third strategy
($R_{4}$ and $R_{4}^\prime$), discussed in the text, can be used to compute the form factor $F_+(q^2)$ 
to a desired precision. Even though our numerical tests are made by using the Wilson quarks, our conclusions 
apply to any lattice QCD action.

\begin{figure}
\begin{center}
\hspace{-1cm}
\resizebox{7.9cm}{5.2cm}{\includegraphics{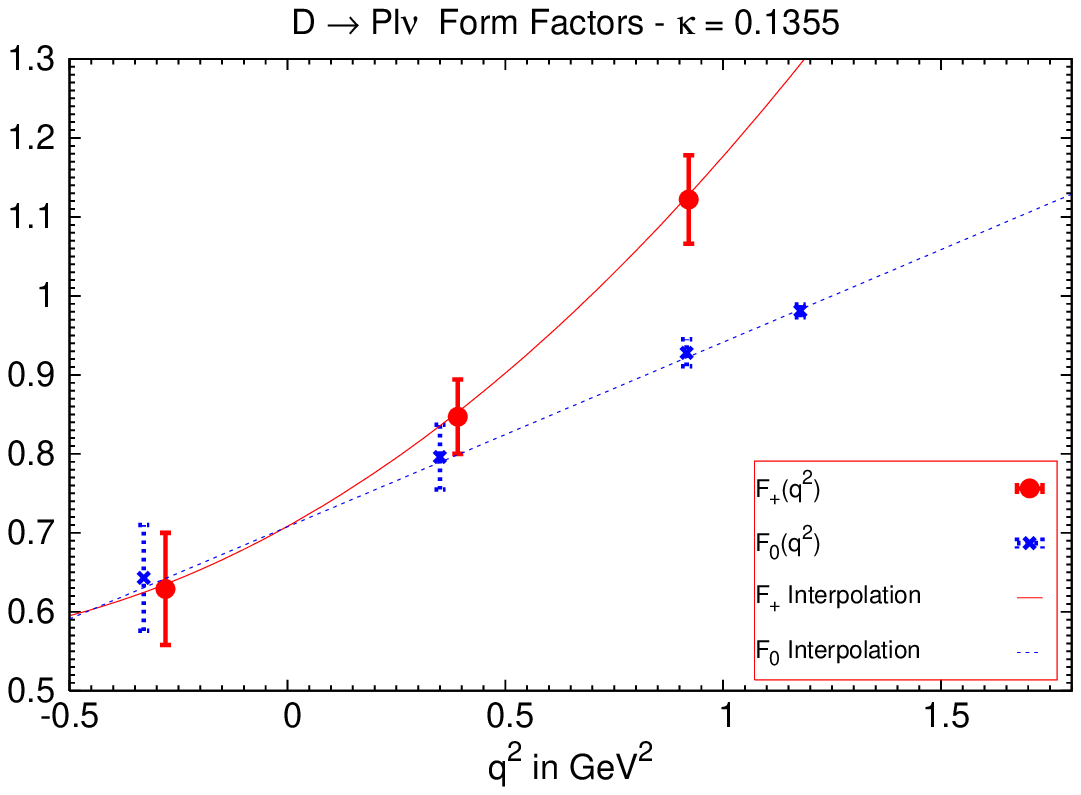}}
\resizebox{7.9cm}{5.2cm}{\includegraphics{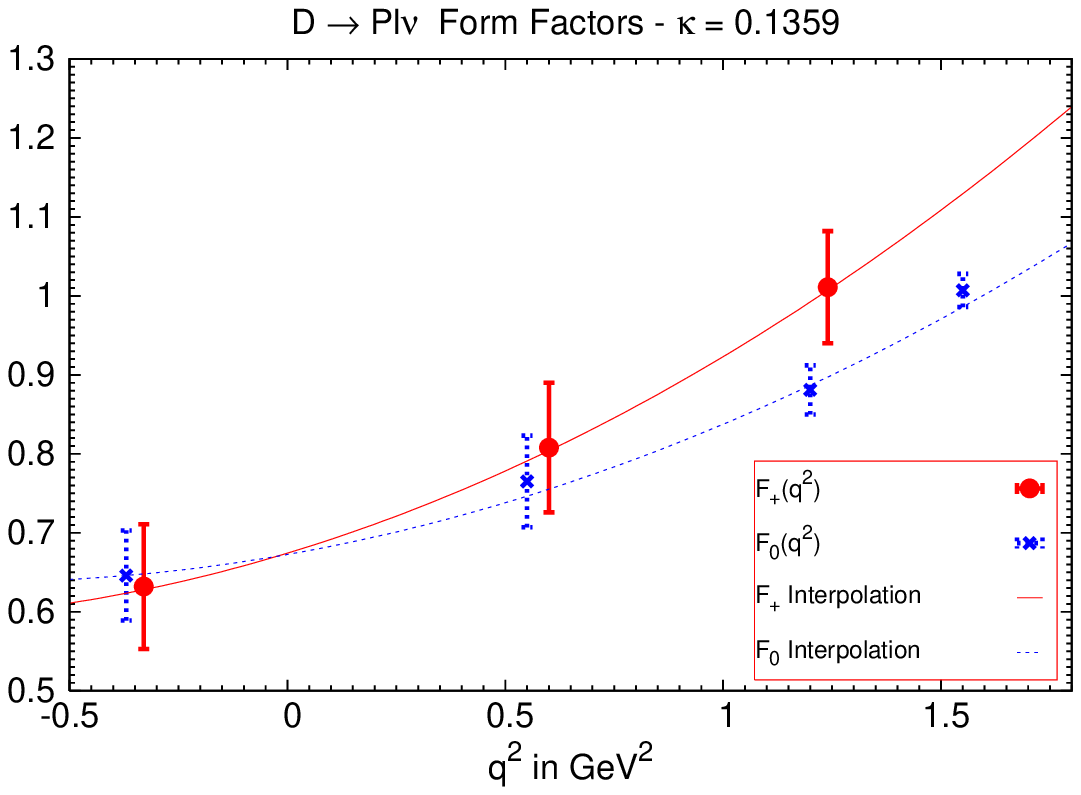}}
\caption{$D\to \pi$ semileptonic form factors for the unphysically heavy pions: 
left panel corresponds to $m_\pi\simeq 750 MeV$, the right one to $m_\pi\simeq 600 MeV$}
\label{fig_ff}
\end{center}
\end{figure}

In fig.~\ref{fig_ff}, we illustrate the results obtained by employing the first strategy at 
three values of the twisting angle $\vec{\theta}$. 
Those will be improved and the results discussed in our forthcoming paper.
\section{Acknowledgments}
We thank the QCDSF collaboration for providing us with their gauge field configurations and 
the \textit{Centre de Calcul de l'IN2P3 \`a Lyon}, for giving us access to their computing facilities.


\begin{thebibliography}{1}

\bibitem{Hashimoto:1999yp}
S.~Hashimoto {\it et al.},
``Lattice {QCD} calculation of $B\to D l\bar \nu$ decay form factors  at
zero recoil,''  Phys.\ Rev.\  D {\bf 61} (2000) 014502.

\bibitem{spqr}
  D.~Becirevic {\it et al.},
``The $K \to\pi$ vector form factor at zero momentum 
transfer on the lattice,''  Nucl.\ Phys.\  B {\bf 705} (2005) 339.

\bibitem{Bedaque:2004kc}
  P.~F.~Bedaque,
``Aharonov-Bohm effect and nucleon nucleon phase shifts on the lattice,''
  Phys.\ Lett.\  B {\bf 593} (2004) 82; 
G.~M.~de Divitiis  {\it et al.},
``On the discretization of physical momenta in lattice QCD,''
  Phys.\ Lett.\  B {\bf 595} (2004) 408.


\bibitem{jlqcd}
N.~Tsutsui {\it et al.}  [JLQCD Collaboration],
  ``Kaon semileptonic decay form factors in two-flavor QCD,''
  PoS {\bf LAT2005} (2006) 357.


\bibitem{Brommel:2006ww}
  D.~Brommel {\it et al.}  [QCDSF/UKQCD Collaboration],
``The pion form factor from lattice QCD with two dynamical flavours,''
  Eur.\ Phys.\ J.\  C {\bf 51} (2007) 335.

\bibitem{Boyle:2007wg}
 P.~A.~Boyle {\it et al.}, 
``Hadronic form factors in lattice QCD at small and vanishing momentum
transfer,''
  JHEP {\bf 0705} (2007) 016.

\end{thebibliography}
\end{document}